# Decent Work Standards and Non-Conventional Work: Measurement Gaps and Uneven Capability Formation in Kenya’s Online Platform Labour Markets

**Farhan Latif, PhD**

United Nations University Institute Macau, Macau SAR, China

latif@unu.edu

## Abstract

Online freelancing through digital platforms is a precarious but sought-after livelihood pathway in contexts with weaker conventional labour markets. This article examines how worker-level decent-work assessments intersect with livelihood strategies and outcomes among experienced online freelancers in Kenya. A convergent mixed-methods study combines survey indicators from 52 workers with interviews with 20 participants. Findings show relatively favourable assessments of work-life balance, inclusion, safety and skill development, alongside weak income consistency, protection and voice. Interviews reveal that measured conditions differ in durability, whether they are institutionally provided or produced through workers’ unpaid effort, and their convertibility into livelihood security and agency. The article argues that decent work provides a necessary normative baseline, but cross-sectional worker-level operationalisations alone cannot reveal the durability, production or livelihood consequences of measured conditions. A capability-informed livelihood perspective complements decent-work assessment by explaining who can convert platform participation into security, agency and future opportunity.

## 1. Introduction

Work is becoming more digital, global, and less tied to fixed workplaces. At the same time, the one-job-for-life, one-employer model of standard employment is losing both its dominance and its normative force (Kalleberg and Dunn, 2016). These emerging work contexts complicate the evaluation of work quality because many established indicators were developed around conventional employment relationships. This research focuses on one such context, online freelancing through digital labour platforms.

Often described as platform work, it is a type of work ecosystem in which well-educated workers compete in a global digital labour market alongside other individuals and small businesses to bid on projects and secure work of varying durations and scopes (Heeks, 2017b). These projects may span occupational areas such as coding, writing, administrative assistance, transcription, and marketing, as well as a range of tasks that may not fit neatly into established occupational categories.

Engaging in this type of work is a noticeable trend in many parts of the Global South, with estimates of 30 per cent annual growth (Heeks et al., 2021; Kässi et al., 2021). Workers typically perform and deliver this work remotely, while platforms such as Upwork and Fiverr mediate access to clients across borders. Analytically, this arrangement combines remote work, freelance contracting, platform mediation, and participation in cross-border labour markets. These respectively shape spatial and temporal autonomy, contractual insecurity and individualised risk, visibility and rule-setting (Graham et al., 2019), and access to global clients alongside various levels of geographic discrimination (Howson et al., 2023)

Concurrently, the continued growth of platform work within the Global South also underscores the need to understand work quality in relation to its structural conditions, its developmental significance for workers, and the local socio-economic and cultural contexts that ultimately shape it. Work may be global, but its meaning and value are shaped by how the quality of those opportunities intersects with local labour markets, social conditions, and livelihood needs. Hence, similar work conditions may have different consequences for what workers can do and sustain in their everyday lives (Graham et al., 2017).

For measuring the quality of work, the ILO's decent-work framework (ILO, 2008, 2013) provides a necessary normative baseline for assessing these conditions. However, the analytical challenge is how to interpret worker-level measures when work is organised outside standard employment and when similar measured conditions may have different livelihood consequences. The challenge is also the presumed context-neutrality of these standards, which may privilege idealised conceptions of good work over what constitutes a viable and worthwhile livelihood in Global South contexts (D'Cruz, 2017; Scully, 2016). This can create a mismatch between normatively defined work conditions and the livelihood significance those conditions assume in particular contexts.

Some relevant developments do extend the decent work agenda to online freelancing. ILO's Decent Work in the Platform Economy Convention No. 193 (ILO, 2026) extends protections across areas including remuneration, social security, automated management, data protection and deactivation. Initiatives such as the Fairwork Project have also adapted labour benchmarks to platform contexts, particularly around pay, contracts, safety and representation (Graham et al., 2020; Heeks et al., 2021; ILO, 2026). These developments strengthen the normative basis for platform-specific protection. However, their primary focus remains on conditions within the platform-work arrangement rather than on how those conditions interact with workers' local labour markets, household circumstances and wider livelihood strategies. They therefore do not resolve whether improved conditions are durable in practice or can be converted into livelihood security and agency under unequal socioeconomic constraints. As a result, worker-level applications of decent-work dimensions provide important but partial accounts when treated as self-sufficient assessments of work and livelihood quality.

This tension forms the starting point for the present study as it problematises what worker-level measures reveal and what they cannot explain about the durability, production, and livelihood consequences of measured conditions. Particularly when work is assessed through the significance workers attach to it within their broader livelihood context, the development interventions and policy changes most relevant to workers can be identified more clearly.

Considering the above, this research moves beyond formal indicators to examine how work conditions intersect with the end goal of work, the impact of work on workers' broader life aspirations, individual capabilities, and long-term sustainability of their livelihoods (Bonvin, 2006, 2012; Klink et al., 2016; Nussbaum, 2005; Sen, 1993, 2001). Following Latif (2024), livelihood sustainability is understood here as platform workers' capacity to convert on- and off-platform work opportunities into income security, skill formation, resilience, and future livelihood options within specific local constraints.

The research uses Kenya as a case study, where there is a growing interest in platform-based freelancing (Melia, 2020). Multiple global and local socioeconomic factors are driving this trend. First is the persistent structural mismatch between job seekers' skills and the local labour market's capacity to absorb skills (The World Bank, 2024). Despite robust economic growth, higher unemployment and persistent job instability in Kenya have made freelancing through digital platforms an attractive alternative. Consequently, a young, educated workforce gravitates towards digital freelancing gigs, attracted not only by financial incentives but also by other benefits, such as greater perceived flexibility in working hours, global connectivity, and better working conditions as compared to the conditions of conventional local jobs (Graham et al., 2017; Lehdonvirta et al., 2021). The analysis is guided by two interconnected research questions.

RQ1: What patterns of work conditions are revealed by worker-level indicators mapped onto selected ILO Decent Work Substantive Elements among experienced online freelancers in Kenya?

RQ2: How are these measured conditions produced, sustained, and converted into livelihood outcomes under differing socioeconomic constraints?

This article contributes to sociological debates on work quality in platform labour by showing that variation lies not only in workers' measured working conditions, but also in their capacity to sustain and convert those conditions into livelihood security and agency. It does so by advancing understanding of the relationship between decent-work standards and the developmental realities of Global South workers (Ayentimi et al., 2022; D'Cruz, 2017; Heeks et al., 2021); linking a capabilities-informed livelihood sustainability perspective with decent work (Austin, 2018; Bonvin, 2012; Burchell et al., 2014; Deneulin and McGregor, 2010; Klink et al., 2016; Monteith and Giesbert, 2017); and highlighting the broader developmental implications of digital freelancing in the Global South (Anwar and Graham, 2022; Latif, 2024; Melia, 2020).

## 2. Decent Work Agenda and Platform Work

Though often considered novel, platform work can be understood as part of a broader shift toward more precarious and insecure labour arrangements, a trend that raises concern over the erosion of labour protections and employment quality (Srnicek, 2017; Standing, 2017). Within this wider context, the International Labour Organization's Decent Work Agenda (DWA) is a relevant normative agenda that aims to safeguard labour standards and advocate for better quality of work for all. At its core, the notion of decent work here refers to work that comes with fair pay, reasonable stability, dignity, voice, and opportunities for personal and professional development (Ahmed, 2003; Ferraro et al., 2016). The DWA is organised around four broad objectives: protecting fundamental rights at work, promoting quality employment and equality of opportunity, extending social protection, and strengthening social dialogue to improve working conditions over time (ILO, 2008).

The ILO's measurement framework for DWA, the decent work substantive elements (SEs), translates the DWA objectives into statistical and legal indicators, primarily for monitoring progress at the country level. The eleven SEs capture multidimensional work characteristics, including employment opportunities, adequate earnings, working time, work–life balance, stability, equality, safety, social protection, and social dialogue (ILO, 2008, 2013). Appendix 1 outlines the selected substantive elements and their relevance to online platform work. Worker-level evaluations of decent work often use survey instruments that link workers' perceptions of their conditions to broader normative indicators of work quality (Duffy et al., 2017; Ferraro et al., 2018; Webster et al., 2015). The decent work framing, therefore, provides a normative baseline for assessing whether work enables workers to participate in the labour market on normatively acceptable terms.

However, the breadth of decent-work indicators makes them adaptable across settings but also creates challenges of interpretation and operationalisation (Bletsas and Charlesworth, 2013; Di Ruggiero et al., 2014). For instance, snapshot worker-level measures may not show whether observed conditions are stable, institutionally guaranteed, or sustained through workers' own unpaid and adaptive labour, particularly in the emerging forms of work (Deranty and MacMillan, 2012; Ferraro et al., 2016). Therefore, relying solely on decent work instruments can miss key aspects of work quality, particularly when work is organised in ways that deviate from conventional arrangements.

### 2.1. *Measuring Decent Work Conditions in Platform Work*

The existing literature on the intersection of platform work and decent work points to three recurring fault lines. First, income volatility, often driven by irregular task flow, opaque pricing, and unstable access to opportunities, remains a persistent weakness in platform labour markets (Fırat et al., 2024; Mokofe, 2022). Second, hidden unpaid labour, including time spent searching for work, preparing bids, screening clients, and maintaining visibility or ratings, erodes both earnings and worker well-being (Ayentimi et al., 2022; Heeks, 2017a). Third, a weak collective voice and limited institutional safeguards leave workers vulnerable (Ferraro et al., 2018; Morales-Muñoz and Roca, 2022).

However, much of the existing research stops at identifying broad decent-work deficits, leaving underexamined how workers navigate these conditions and convert them into opportunities or constraints within their wider lives. This deficit-based approach creates an important explanatory gap between documenting the objective conditions of platform work and understanding their implications for worker well-being and livelihood development (Bonvin, 2012). When these decent work indicators are treated as proxies for workers' wider well-being, such deficit-based assessments may therefore imply a more direct and uniform relationship between formal work conditions and worker well-being than the empirical realities of platform labour support.

Moreover, differences in work quality among platform workers may reflect structural inequalities but also variations in qualifications, skills, and experience, given the sector's low formal entry requirements (Latif, 2024). In platform work, workers' skills, socioeconomic positions, motivations, and sociocultural backgrounds produce uneven configurations of opportunity and constraint. This points to an analytical disconnect between assessing the work and understanding the worker, whose circumstances and conversion factors shape whether and how particular working conditions influence individual development priorities and enable valued livelihood outcomes.

### *2.2. Capabilities as the Foundation for Livelihood Sustainability*

Favourable decent work evaluations may not necessarily establish that the underlying conditions are secure or that they expand workers' substantive freedom (Bonvin, 2012). A freelance worker may report adequate income or autonomy at one moment while remaining unable to refuse undesirable work, withstand a period without contracts, or convert platform experience into wider livelihood opportunities. Some jobs may offer good income and satisfy formal standards yet remain exhausting or weakly connected to a worker's life aspirations and well-being (Ragadu and Rothmann, 2023; Sen, 2001).

The capability approach therefore complements decent work assessment by asking what workers can genuinely do and become through the conditions they experience (Bonvin, 2012; Nussbaum, 2005; Sen, 1993). In the capability approach, well-being is evaluated through people's genuine opportunities and substantive freedom to achieve valued functionings, such as building skills, improving social capital, or being healthy (Sen, 1993, 2001). This approach shifts attention from whether work appears acceptable on formal indicators to what workers can genuinely do and become through it. A key distinction, relevant to the present analysis, is between functionings and capabilities. Functionings are achieved states and activities, such as earning an income, acquiring skills, or maintaining health, while capabilities refer to the real freedom to achieve valued outcomes and choose among meaningful alternatives (Sen, 1993, 2001). Hence, work could generate visible achievements without expanding substantive freedom. For instance, a worker may earn decent income while lacking the ability to refuse underpaid work, leave an exploitative arrangement, or use acquired experience to pursue broader life aspirations. Decent work indicators can identify important conditions of work, but they may not necessarily reveal the wider choice set available to workers.

This approach also considers contextual circumstances through the notion of conversion factors. Conversion factors are the personal, social, and environmental conditions that shape what people can do with the resources and opportunities available to them (Deneulin and McGregor, 2010; Sen, 2001). Resources and reported satisfaction matter, but neither establishes the extent of workers' substantive freedom. The same platform income, skill, or rating may have different livelihood value depending on a worker's cost of living, family obligations, access to institutional protection, and capacity to save. Conversion factors therefore help explain why the same work-related resource can have different livelihood value across workers.

This consideration is particularly important in Kenya's platform economy, where workers often evaluate opportunities against highly constrained alternatives and challenging socioeconomic conditions (Melia, 2020). That said, Sen cautions against relying too heavily on subjective satisfaction because conversion factors and constrained alternatives can shape adaptive preferences (Sen, 1982). Reported satisfaction may therefore indicate only that platform work is preferable to the available options, without establishing that its conditions are adequate. Subjective evaluations and decent work standards consequently serve complementary purposes: subjective accounts can show how conversion factors shape the value workers derive from platform work, while decent work standards provide an external benchmark for assessing whether its conditions are adequate.

#### 2.2.1 Capability for Work and Workers' Voice

Bonvin (2006, 2012) identifies two conceptions of capabilities particularly relevant to the relationship between work and livelihood sustainability: capability for work and capability for voice. Capability for work is the substantive freedom to choose, refuse, and shape one's employment. It depends on opportunity freedom, supportive conversion factors within the worker's local environment, and a viable exit from work that does not support valued functionings (Bonvin, 2006, 2012). In online freelancing, capability for work operates across two interconnected labour markets: workers' freedom to refuse, shape, or leave undesirable platform work, and how access to platform work reshapes these freedoms in relation to conventional employment.

Capability for voice concerns workers' substantive freedom to influence the rules governing their work (Bonvin, 2006, 2012). For platform workers, this may include influencing platform rules, negotiating contract terms, and seeking effective dispute resolution. From this perspective, collective action could strengthen capability for voice but is not equivalent to it. Unions, professional associations, and other forms of collective organisation may provide resources through which platform workers may exercise voice. But the concept ultimately concerns whether workers can express their preferences and have them considered in decisions affecting their work, regardless of the institutional form through which this occurs.

## 3. Methodology

The article utilises a convergent mixed-methods approach. The quantitative component provides a cross-sectional description of selected worker-reported conditions. The qualitative component examines how those conditions are produced, whether workers experience them as durable, and how they are converted into livelihood capabilities. Data for both components were collected and analysed during the same period, and the quantitative and qualitative findings were integrated during interpretation to identify areas of convergence, divergence, and complementarity (Creswell and Plano Clark, 2011). Quantitative findings are presented in the results section, whereas qualitative findings are anchored in the results and integrated into the discussion to preserve the nuance of participants' perspectives. The research received ethics approval through a university ethics board.

### *3.1. Quantitative Segment*

The quantitative component used an online questionnaire containing 38 closed-ended items adapted from established decent-work measures and mapped onto nine of the eleven substantive elements (Duffy et al., 2017; Ferraro et al., 2018; ILO, 2008, 2013). The questionnaire was administered through the Qualtrics platform to 52 experienced online freelancers in Kenya between January and July 2024. Most attitudinal items used five-point response scales ranging from 0 to 4, while a small number used binary or continuous response formats.

Given the modest, non-probability sample and the multidimensional nature of the ILO substantive elements, the quantitative analysis focused on descriptive benchmarking rather than psychometric scale development or population inference. Within each substantive element, items were combined only when they addressed the same narrowly defined condition and used compatible response formats. Conceptual coherence was the primary basis for aggregation, and conceptually distinct items were treated as single-item indicators. Composite scores were calculated as the mean of their constituent items after coding all items in a consistent direction, preserving the original response metric. Appendix 1 documents the number of constituent items, response formats, aggregation rules, internal-association statistics, and interpretation of each indicator.

### 3.2. Qualitative Segment

Semi-structured interviews were conducted with 20 survey participants (12 female, 8 male) between January and July 2024. Interviews were conducted in person, lasted about 1 hour, were in English, and were audio-recorded with consent and transcribed. The interviews were analysed thematically (Clarke and Braun, 2017), combining deductive attention to the decent-work domains with inductive identification of recurring explanations of how measured conditions were produced, sustained, or converted into livelihood outcomes. At the interpretive stage, capability concepts were used to examine differences in agency, conversion factors, and capability for work and voice, rather than as a fixed coding template. Narrative attention was retained where participants' accounts depended on changes over time or sequences of work and livelihood decisions (Riessman, 2008). The study integrated survey and interview findings by examining convergence, divergence, and complementarity around the measured decent-work conditions. Quotations are lightly edited for readability, and participants are identified by pseudonyms.

### 3.3. Research Participants

For this research, participants were recruited through a combination of purposive, voluntary non-probability sampling strategies. This included posting calls for participation in online freelancer communities on social media and Upwork, connecting with individuals affiliated with digital skills training programmes in Kenya, engaging attendees at freelancer networking events, and meeting workers at co-working hubs in Nairobi, Kenya.

Recruitment channels significantly shape platform-work samples and researcher perspectives (Eurofound, 2018; OECD et al., 2023). This methodological challenge was evident during recruitment. When workers encountered the study through the same digital channels they used to access paid work, they often interpreted participation as another form of labour. This risked making participation more transactional and less candid, while also attracting mainly less experienced workers or those struggling to secure sufficient work. More experienced workers with stable or higher-value contracts were less interested in platform-driven recruitment because participation offered little practical value relative to their existing work opportunities.

Research participants are primarily from across the Nairobi metropolitan area. Most participants were highly educated, with 80.8% holding a bachelor's degree or higher (Table 1), and all had at least one year of experience in the platform economy. The study therefore represents experienced freelancers rather than novice or prospective platform workers.

Table 1. Research Participants' Demographics

| Demographic Characteristics | Frequency | Percent |
|---|---|---|
| Gender | | |
| Female | 23 | 44.23% |
| Male | 29 | 55.77% |
| Age Group | | |
| 18–24 | 8 | 15.38% |
| 25–34 | 27 | 51.92% |
| 35+ | 17 | 32.69% |
| Education Level | | |
| Some college or lower | 10 | 19.23% |
| Bachelor's or higher | 42 | 80.77% |
| Platform Work Experience | | |
| 1 to less than 3 years | 10 | 19.23% |
| 3 -5 years | 19 | 36.54% |
| More than 5 years | 23 | 44.23% |

On average, participants earned approximately 69% of their total income through platform work. Participants reported working an average of 38.8 paid hours per week, alongside 19.7 hours of unpaid labour, including tasks such as job search, communication, proposal writing, and other unpaid activities common in the platform work economy (Table 2). Additionally, participants invested an average of 16.4 hours per week in unpaid skill development activities.

Table 2. Platform Work Intensity Measures

| Work Hours, Earnings, and Skill Development | Mean | SD | P50 | P75 |
|---|---|---|---|---|
| Platform Income as a Percentage of Total Income | 69.35 | 25.3 | 75 | 90 |
| Paid Work Hours/Week | 38.79 | 18.47 | 39 | 52 |
| Unpaid Work Hours/Week | 19.72 | 21.78 | 10 | 27.5 |
| Skill Upgrading Hours/Week | 16.43 | 17.31 | 10 | 20 |

## 4. Results

The results show uneven decent-work conditions among the sampled freelancers rather than a uniformly positive or negative pattern. Several reported conditions are relatively favourable, particularly work–life satisfaction, perceived inclusion, physical safety, and opportunities for skill development, although the qualitative findings show that these outcomes arise through different aspects of the work arrangement. At the same time, the results point to persistent weaknesses in income predictability, institutional protection, healthcare access after joining platform work, and collective voice. Table 3 summarises the indicator means and standard deviations, indicator construction and internal-association statistics, and qualitative synthesis anchors.

Table 3. Study indicators mapped to decent-work substantive elements and qualitative synthesis anchors

| Decent-work substantive element and study indicator | Mean (SD) | Indicator construction and internal association | Qualitative synthesis anchor |
|---|---|---|---|
| **Employment Opportunities** | | | Platform-mediated access to cross-border clients expands opportunity, but its livelihood value remains shaped by weak local alternatives and dependence on continued platform access. |
| Work Opportunity Access through Platforms | 2.48 (0.72) | Mean of two items; r = 0.48 | |
| Alternative Work Opportunities | 2.38 (0.95) | Single item | |
| Extent of Financial Hardship without Platforms | 2.75 (1.06) | Single item | |
| **Adequate Earnings and Productive Work** | | | Cross-border client access can improve earnings relative to local alternatives, while freelance task volatility and unpaid platform-search and reputation work undermine income consistency. |
| Financial Satisfaction | 2.37 (0.84) | Single item | |
| Income Consistency | 1.30 (0.97) | Single item | |
| Skill Development and Productive Use | 2.60 (0.83) | Mean of three items; α = 0.71 | |
| **Decent Working Time** | | | Remote delivery reduces commuting, while some control over freelance scheduling allows workers to reallocate time across livelihood activities. |
| Schedule Autonomy and Negotiation Power | 2.18 (1.00) | Mean of two items; r = 0.40 | |
| **Combining Work, Family, and Personal Life** | | | Remote presence and some control over task timing reduce forced trade-offs between earning income and sustaining caregiving, learning, family, and social roles. |
| Work–life Satisfaction and Fulfilment | 2.93 (0.70) | Mean of three items; α = 0.80 | |
| **Stability and Security of Work** | | | Perceived growth reflects selective stability built through reputation, repeat clients, and diversification, while platform access remains fragile because regional restrictions can erase accumulated ratings and experience. |
| Stability and Growth of Work | 2.81 (0.73) | Mean of three items; α = 0.76 | |
| **Equal Opportunity and Treatment** | | | High inclusion scores capture respectful day-to-day interactions, but interviews show that Kenyan location, Global South status, and client trust biases still shape who is hired and protected. |
| Perceived Inclusion and Non-Discrimination | 3.68 (0.53) | Mean of two items; r = 0.38 | |
| Platform Policy Protection | 1.12 (1.34) | Single item | |
| **Safe Work Environment** | | | Positive safety and low stress scores partly reflect distance from office politics and local workplace hierarchies. |
| Physical and Mental Well-Being and Safety | 3.07 (0.75) | Mean of two items; r = 0.46 | |
| Stress from Platform Work | 1.54 (0.90) | Single item | |
| **Social Security** | | | |

| Decent-work substantive element and study indicator | Mean (SD) | Indicator construction and internal association | Qualitative synthesis anchor |
|---|---|---|---|
| Perceived Healthcare Access in Traditional Jobs Compared to Platforms (higher is better) | 2.46 (1.02) | Single item | A systemic shift of risk management onto individuals is coupled with a means-to-an-end logic, in which formal benefits are often seen as less important when they do not translate into adequate protection. |
| Perceived Change in Healthcare Access Compared with Traditional Employment (higher is better) | 1.77 (0.98) | Single item | |
| Percentage of Platform Income Saved | 25.25 (20.13) | Single item | |
| **Social Dialogue** | | | Weak formal collective action does not always mean solidarity is absent in platform work; workers may organise and build robust peer networks for information, skills, and support, but these rarely translate into platform-level bargaining, as many workers do not regard platform-level bargaining as a priority. |
| Opportunities for Collective Action | 0.16 (0.32) | Mean of two binary items; r = 0.51 | |
| Labour Organisations' Helpfulness | 0.31 (0.47) | Single binary item | |
| Participation in Collective Action | 0.21 (0.33) | Mean of two binary items; *r* = 0.37 | |
| Public Policy Participation | 0.66 (0.99) | Mean of two items; r = 0.58 | |
| Freedom of Expression | 2.20 (1.10) | Mean of two items; r = 0.40 | |
| Client Attention to Labour Concerns | 1.62 (1.22) | Single item | |
| Participation in Platform Administrative and Policy Decisions | 1.07 (1.18) | Mean of two items; r = 0.71 | |

Note. N = 52. Values are M (SD). Unless noted otherwise, scores range from 0 to 4 where higher scores indicate more of the named condition. Healthcare-change indicators are centred at 2 (no change). Savings range from 0 to 100%. Binary indicators range from 0 to 1, with M representing the mean proportion endorsed. r denotes the inter-item correlation, and α denotes Cronbach's alpha.

Among the higher reported values were perceived inclusion, well-being and safety, work–life satisfaction, and stability and growth. Lower values were reported for income consistency, platform policy protection, healthcare access after joining platform work, and most social-dialogue indicators.

The clearest contrast is between opportunity and security: access to platform work and financial satisfaction were higher than income consistency. Social and interpersonal indicators were generally higher than institutional-protection and voice indicators. Perceived inclusion, physical and mental well-being and safety, and stress from platform work are rated more favourably than platform policy protection, healthcare access after joining platform work, and social dialogue indicators. The lowest scores appear in workers' ability to influence platform administrative and policy decisions, client attention to labour concerns, participation in collective action, and public policy participation.

## 5. Discussion and Qualitative Synthesis

Platform-work research often foregrounds structural precarity and demand-side pressures. In Kenya, however, labour-market insecurity also extends beyond platforms, so workers often manage rather than escape precariousness.

Understanding how workers navigate that precariousness within their socioeconomic constraints is therefore essential for identifying locally viable support and interventions to improve working conditions across diverse contexts.

Because platform work often forms one part of a broader livelihood strategy, workers evaluate its deficits against both employment standards and its contribution to income, options, and security. Measured work quality and lived livelihood sustainability therefore do not always move together. Platform precarity remains important, but workers encounter and manage it from unequal positions. The research findings also show that no single, uniform profile of decent work exists in the Kenyan platform economy. Two broader points are especially important to understanding this fluidity.

First, a favourable score on a worker-level indicator does not automatically establish that the underlying condition is sustainable. The qualitative findings show that some favourable conditions are temporary, depend on continuous worker effort, or are available primarily to workers with sufficient resources and bargaining power. This may take the form of limited room to refuse or renegotiate undesirable gigs, as well as substantial unpaid effort that remains largely uncaptured in decent work standards. The interviews qualify the survey indicators by showing their temporal stability or fragility, whether favourable conditions are institutionally provided or worker-produced, and how conversion factors shape their livelihood value.

Second, decent-work dimensions do not carry the same immediate livelihood value in workers' assessments, because their practical significance is shaped by institutional quality, platform constraints, and the socioeconomic conditions that determine whether they can be converted into valued outcomes. This study examines these contradictions through five empirically grounded themes that emerged from the qualitative analysis.

### 5.1. Platform Work as an Adequacy Bridge under Constrained Off-Platform Opportunities

Kenyan freelancers experience platform work as an adequacy bridge, an incomplete and unstable form of work that nevertheless helps them to manage weak local employment, income pressure, and limited upward mobility. Platform work is therefore preferable to local alternatives, though it is not normatively adequate. As one participant explained:

> "With the office work, yeah, it is guaranteed [monthly income], but it is not that guaranteed. You can get fired during the week or maybe during the month. The same way it is on the platforms" (Clara).

Stable employment relationships with good benefits and adequate income are not readily available in many labour markets in the Global South (Scully, 2016). When both on- and off-platform opportunities are precarious, the relatively higher financial satisfaction associated with platform-based income is a decisive factor for some workers, offsetting instability deficits through perceived earnings adequacy.

> "I cannot be working [in traditional jobs] for 10 hours a day and [they are] giving me 40,000 Kenyan Shillings. My house rent is 25,000. I have other bills: electricity, water, and garbage. What am I left with? I have no money for food, for transport, for taking care of myself, my parents, my son, and the education for him. It is not worth it" (Clara).

Participants often linked their relative financial satisfaction with platform work to the weaker conditions they had experienced or perceived in locally available employment. Limited local opportunities to valorise skills acted as an adverse conversion factor, pushing some workers toward global digital piecework despite weak contractual security and inconsistent income. Platform work could therefore improve immediate economic prospects without resolving the insecurity shaping workers' choices. Immediate opportunity freedom can therefore be lucrative while remaining insufficient to counter long-term economic insecurity.

> "It is tricky. In December last year, I did not work; there was no work. This January, a lot of work has come. Then the challenge comes when work is too much. Maybe you have sent proposals, and three or four clients respond at the same time, all wanting their work done within a month or so" (Theo).

Unpredictable demand converts this economic reliance into heightened risk, eroding key pillars of career stability, predictable earnings, manageable workloads, and the capacity to plan financially. These are well-documented decent work deficits in the global platform economy (Anwar and Graham, 2021; Wood and Lehdonvirta, 2021). However, a key livelihood sustainability challenge for many Global South workers remains the shared constraint on capability for work across both conventional and platform settings.

> "With the office work, yeah, it is guaranteed [monthly income], but it is not that guaranteed. You can get fired during the week or maybe during the month. The fact that somebody hired you to be in the office does not mean that you are there forever; it is your performance. The same way it is on the platforms" (Clara).

This instability in both on-platform and off-platform work conditions leads workers to conceptualise employment as a fundamentally precarious and unpredictable undertaking, with platforms bridging the gaps. This insecurity encouraged workers to combine platform work with other income-generating activities.

### *5.2. The Contextual Value of Temporal Autonomy as a Decent Work Condition*

Flexibility in work timing and remote delivery gave some workers greater room to combine paid work with other livelihood and domestic activities. As a result, temporal autonomy could expand the preferred activities workers can sustain alongside income-earning.

> "As much as I wanted to work in a traditional office job, I usually have other things I am doing on the side; at that time, I was freelancing and also baking cakes for sale. So, I wanted the flexibility of being able to deliver cakes" (Triza).

For some workers, remote delivery eliminates substantial commuting time in Nairobi traffic and allows them to remain physically present for household and livelihood activities. Separately, some control over when freelance tasks are completed reduces the constraints of fixed schedules. The collective outcome is often reduced time poverty, which broadens the choice set and gives some freedom to invest in multiple forms of productive socioeconomic activity without being locked into a rigid employment routine of long hours, lengthy commutes, and inadequate pay. In this context, workers are conditioned to see livelihood success not as linear progression within a single decent job (which remains unattainable for many), but as the freedom to navigate across activities that collectively sustain well-being, demonstrating how temporal autonomy can serve as a foundation for adaptive and sustainable livelihoods in low-resource contexts.

> "I work according to my schedule. I keep my friendships; family, for me, is very important. I do not miss out on family events. Even if it means I am taking time out, when it comes to my daughter's events, I have attended all of them" (Rebecca).

Time and income are often competing and scarce resources in almost all regions. However, in low-resource environments, this competition is particularly evident, meaning workers must frequently choose between earning activities and investing in other central functionings, such as caregiving, learning, rest, or civic participation. These trade-offs are also gendered: participants' accounts suggest that control over time can have particular livelihood value for women carrying disproportionate unpaid-care responsibilities.

> "It is life-changing. As a woman and a mother with two children, working remotely has given me so much time to be able to do the things I need to do as a mother while I'm still working. I don't have to give up on either of those things. I don't know if I can describe the impact. It's worth the risk of not having, like, what can I say... guaranteed income." (Mercy)

Even a minor increase in temporal agency and autonomy can be developmentally significant, as it could allow workers to achieve higher-order capabilities without sacrificing income. In contexts where social support systems such as subsidised childcare, public transport, or health coverage are weak, the opportunity cost of time becomes higher, and the freedom to pursue more subjectively meaningful or life-enriching activities narrows significantly.

### *5.3. Interpersonal Civility, Structural Exclusion, and the Limits of Surface-Level Equality Readings*

A favourable score on the study's inclusion and non-discrimination indicator captures reported interpersonal treatment but does not exhaust the broader equality dimension. The interviews support the view that many workers experience day-to-day platform interactions as more civil and sometimes less hostile than some local workplaces.

> "You don't have to deal with office politics, the backbiting. These things don't happen online. You have that peace of mind" (Clara).

For some workers, especially those coming from hierarchical or exclusionary workplace settings, remote work can feel less visibly oppressive and more manageable in everyday interaction (D'Cruz and Noronha, 2016; Malik et al., 2017). At the level captured by the survey indicators, the interpersonal dimension of equality and safety appears to be met to a satisfactory extent. However, interviews also suggest that this day-to-day environment of non-discrimination and respect is not granted, but rather self-constructed after intentionally filtering out the prejudice and discrimination in the platform work economy.

> "If I change my profile, I retain the same name, everything same, same skillset, same quality, same ratings, but I change my location to the US, you definitely see a better quality of jobs appear" (Elijah).

The questionnaire indicators therefore measure everyday civility more effectively than geographic discrimination in access to clients and assignments. Kenyan workers are aware that many clients harbour biases. Although platforms are expected to reduce information and trust barriers (Kenney and Zysman, 2016), geographic hierarchies can remain embedded in how workers are seen, trusted, and hired (Srnicek, 2017).

> "Obviously, they discriminate by the geographical regions, that people who will not trust that someone in a third-world country has the skills to do the job. They will ask for a sample, and then once you submit the sample, they come back and say, oh, I am very surprised you are able to do this. I did not think that people from your country could do this" (Mercy).

The positive experiences of civility are partly sustained through worker adaptation. Some workers described avoiding clients whose prior behaviour signalled likely prejudice or low trust. In this sense, non-discrimination in day-to-day experience is sometimes achieved through anticipatory filtering rather than guaranteed by the platform environment itself. Workers learn where not to apply, how to present themselves, and which interactions are less likely to be worth the emotional or economic cost. For workers with enough choice to avoid undesirable clients, the result is a more selective, curated experience of civility that can mask the broader exclusion shaping access to opportunities.

In the platform economy, civility can therefore coexist with structural exclusion, and the latter may remain partly hidden precisely because workers have learned to navigate it strategically. While decent work indicators can indicate whether workers feel respected or safe in interactions, they may be less well suited to showing how workers remain positioned within unequal labour hierarchies that shape who gets seen, trusted, and hired in the first place.

### *5.4. Protection, Benefits, and the Limits of Institutional Support in Assessing Decent Work*

The survey results on social protection, healthcare access, and savings point to different limits. Online freelancing through platforms generally lacks social benefits and safety nets such as healthcare, retirement support, and other essential provisions, including vacation pay, paid sick leave, and maternity leave (Ayentimi et al., 2022; Graham et al., 2019; Heeks, 2017a; Mokofe, 2022). These are critical limitations and represent key deficits in decent work within the platform economy. However, from a capabilities perspective, Kenyan platform workers present a compelling challenge to conventional indicators of job quality. Standard labour frameworks often treat employment benefits, such as health insurance, retirement contributions, or paid leave, as central markers of decent work (ILO, 2008). However, these benefits can be of limited value without adequate income.

> "If I have remote work that pays me a good amount of money, I will consider saving some money and putting it in my pension or savings account, rather than having an on-site job that pays very little and gives me the pension benefits, which are [often] uncertain" (Mable).

Workers do feel excluded from these benefit schemes, a key deficit in the platform economy; however, they emphasise the importance of immediate financial control and prefer higher earnings that they can manage independently, whether to pay out of pocket for private healthcare beyond what traditional benefits offer in Kenya, save for retirement, and create a buffer against emergencies. Their focus is naturally on what substantive protection the entire employment arrangement enables them to achieve. The value of employment benefits cannot be evaluated independently of the institutional regime through which they are delivered.

> "What I learned is, when you are getting a job or a client, negotiate a higher pay so you can cover some of those benefits for yourself, like health insurance" (Triza).

The implication is not that employment benefits are not desired, but that benefit-centred measures can mistake such an institutional target for improved security. The benefits attached to low-paid or unreliable jobs may offer protection but still fail to enhance workers' capacity to absorb shocks, invest, and sustain valued livelihood choices. Moreover, workers also noted that benefit schemes in some conventional employment often do not provide the private care needed when public hospitals are inadequate.

> "Money answers everything. Last year was my best year. I made enough money to buy a plot without taking a loan. If I was still in an office job, I wouldn't have had that freedom. I wouldn't have been able to pay for that plot outright. Instead, I'd have been forced to go to my SACCO, take a loan, and still be repaying it right now" (Clara).

Workers have a simple yet easily overlooked means-to-an-end logic: they evaluate work benefits according to the protection and freedoms those benefits actually enable. Some Kenyan freelancers therefore assess earnings and benefits as an interdependent employment package, prioritising sufficient income and the capacity to save where formal benefit schemes are perceived as weak, uncertain, or inadequate for accessing the care they need. The survey finding that participants saved, on average, about one-quarter of their platform income illustrates this individualised approach to managing livelihood risks.

Work evaluation should therefore ask not only whether benefits are formally present, but also how these entitlements intersect with other work quality indicators and whether they expand workers' genuine freedoms. That said, understanding how platforms shift the burden of protection onto workers is essential to offering suitable alternatives. This shifted burden of responsibility and the individualisation of risk also appear across several decent work dimensions in this study, making individual approaches the deciding factor between those who can cultivate decent working conditions for themselves and those who cannot. While this may enhance agency for workers who can afford to be selective, refuse poorly paid gigs, and negotiate compensation sufficient to offset missing protections, it deepens inequality for those whose immediate dependence on platform income leaves them little freedom to reject undesirable work or bargain for better terms.

### *5.5. Worker Voice, Informal Solidarity, and the Limits of Formal Social Dialogue*

The very weak social dialogue scores in the survey mark a clear decent work deficit. Yet workers' limited prioritisation of formal collective action also points to a deeper mismatch between conventional ideas of worker voice and the ways many platform workers pursue livelihood security and think about the value and purpose of solidarity. The first challenge is that the platform labour market's nature creates constant competition among geographically dispersed workers, many of whom may never interact, making traditional forms of worker solidarity and collective action less immediately feasible or useful.

> "[On the subject of unionisation and collective action] it is something that I am not looking for" (Marvin).

Collective mobilisation requires not only time but also significant cognitive resources, including emotional investment, political awareness, and coordination efforts (Bonvin, 2012; Heckscher and Carré, 2006). Because more favourable work terms are often accessible to highly rated workers, their incentive to engage in collective action is diminished. As a result, many workers do not see labour organisations as helpful in enhancing their voice.

Kenyan freelancers' accounts and the presence of many collaborative networks suggest that solidarity is not absent but is organised around livelihood navigation rather than oppositional mobilisation. It appears through social media groups, peer communities, skills-sharing spaces, and in-person networking events that help workers find opportunities, learn platform strategies, build confidence, and remain socially connected within an otherwise individualised labour market.

> "I am a member of one community; we are trying to connect as freelancers and share skills. These communities are growing into organisations where we are now trying to teach skills to other youth so that they can also join these online opportunities" (Theo).

The purpose of such solidarity is therefore less to challenge platforms directly than to make platform work more usable in a constrained local economy, where access to information, tacit knowledge, referrals, and peer encouragement become part of workers' livelihood infrastructure. The weak social-dialogue scores therefore reflect limited formal worker organisation alongside a deeper tension between conventional models of collective action and the livelihood strategies through which workers navigate platform markets.

In globally distributed freelance markets, workers compete through individualised reputation, client access, pricing, speed, and skill differentiation, making personal market positioning the main route to livelihood sustainability. Under these conditions, the suitability of conventional forms of institutionalised collective action is not self-evident, because workers, particularly in the Global South, do not occupy a unified position within the labour market. Their interests are divided not only by income, skill, gender, ethnicity, and class, but also by global hierarchies between North and South and by inequalities within the Global South itself. For workers, whose platform access is already fragile, stronger regulation, higher standards, or formal protections may appear desirable in principle but risky in practice if they are perceived to make the market more attractive to better-resourced competitors. Social dialogue in platform work, therefore, cannot simply assume that conventional employment-based models of representation would genuinely improve workers' capacity for voice. It must address the more basic question of how workers can gain influence over platform conditions without losing the access, flexibility, and individualised livelihood strategies through which they currently convert platform work into security, mobility, and future capability.

## 6. Conclusions

The challenge of assessing decent work in online platform labour lies not only in what worker-level measures can capture, but also in how the conditions they identify are interpreted in relation to workers' wider livelihood circumstances and priorities. As this study showed, in Kenya's online freelance economy, work is often valued less as a complete employment arrangement than as a partial and unstable, yet strategically important, component of livelihood sustainability. For some workers, it can provide relatively lucrative opportunities, greater autonomy, and a sense of empowerment compared with locally available alternatives even when it does not score well on decent work appraisals. These positives do not eliminate deficiencies in income security, protection, voice, and working conditions; nor do such deficiencies render the livelihood value of platform work inconsequential. Platform work, therefore, may fall short of several decent-work standards while still expanding workers' practical room for manoeuvre. Conversely, positive assessments may represent genuine improvements in workers' lives while also reflecting adaptation to constrained alternatives rather than the secure formation of capabilities. This tension complicates both celebratory and deficit-based accounts of platform work.

Decent-work standards provide a necessary normative floor for identifying deficiencies in employment quality. However, assessing working conditions alone cannot fully explain why similar conditions produce different consequences for different workers. A worker-centred capability lens adds an interpretive layer by showing how skills, socioeconomic position, household responsibilities, and other conversion factors shape what workers can derive from platform participation. It thereby adds an interpretive layer to decent-work assessment, connecting employment conditions with the processes through which workers convert platform access into livelihood security, agency, and future opportunities. The contribution is consequently a more context-sensitive account of how working conditions become consequential within workers' wider lives.

Where governments, NGOs, and local institutions have limited leverage over global platforms, interventions should also strengthen the local conditions through which workers convert platform access into more durable capabilities. This includes attention to household constraints, skill trajectories, multiple income sources, reputation-building, saving, and other strategies workers use to manage insecurity. Such support should complement, not substitute for, improvements in platform governance, protection, and voice. The objective should be to strengthen the conditions under which platform access can be converted into durable capabilities, while continuing to pursue improvements in platform governance, worker protection, and collective voice. Supporting workers' adaptive strategies should therefore complement, rather than substitute for, efforts to address the structural conditions producing insecurity.

These conclusions must be considered in light of the study's relatively small, Nairobi-centred sample and its concentration on workers' experiences during a particular period of their platform careers. Longitudinal research is needed to trace movement between supplementary, transitional, and primary dependence on platform income and to determine when temporary gains develop into more durable livelihood capabilities. Further research should also distinguish the effects of digitally enabled remote work from those arising from freelance instability and examine how gender, household position, occupational differences, and sociocultural context shape workers' experiences and outcomes. Ultimately, assessing platform work in development contexts requires asking not only whether it meets established standards or creates new opportunities, but for whom those opportunities are valuable, under what circumstances, and whether workers can convert them into livelihoods they have reason to value.